\documentclass[twocolumn]{aastex63}

\usepackage{rotating}
\usepackage{graphicx}
\usepackage{amssymb}
\usepackage{amsmath}
\usepackage{multirow}
\usepackage{hyperref}
\usepackage{lineno}
\usepackage{lipsum}
\usepackage{gensymb}
\usepackage{todonotes}
\usepackage{float} 

\revised{\today}

\def\g-rays{{$\gamma$ rays}}

\def\aap{A\&A}

\def\apjl{ApJL}
\def\apjs{ApJS}

\newcommand{\R}[1]{#1}
\begin{document}
\title{\texttt{FermiPhased}: A tool for phase-resolved likelihood analysis of Fermi-LAT data}

\author[0000-0003-3540-2870]{Alexander Lange}
\affiliation{Department of Physics, The George Washington University, 725 21st Street NW, Washington, DC 20052, USA.}
\author{B.B. K.}
\affiliation{Arlington, VA 22204, USA.}

\email{alexlange@gwu.edu}

\begin{abstract}
The Fermi Large Area Telescope has enabled detailed studies of high-energy astrophysical sources. To support analysis, we present FermiPhased, a flexible, open-source tool for phase-resolved studies of pulsars, binaries, and other periodically variable sources. Built on the Fermipy framework, FermiPhased offers three modes: standard, adaptive (fixed counts), and joint phase-resolved analysis, enabling users to flexibly bin data based on phase, count statistics, or jointly fit different epochs of interest. FermiPhased is optimized for parallel execution and use on computing clusters. It enables parallelized extraction of phase-resolved fluxes, spectra, and intermediary data products, with tutorials and documentation available on GitHub.
\end{abstract}

\section{Introduction}
Since the Large Area Telescope (LAT) launched in 2008 aboard the Fermi Gamma-ray Space Telescope mission, countless variable gamma-ray sources have been discovered. Designed to detect photons between 20\,MeV and roughly 1\,TeV, Fermi--LAT continuously scans the sky roughly every three hours, providing an all-sky survey of the high-energy universe \citep{atwood2009}. The past 17\,years have provided a unique opportunity to study known pulsars, and high and low-mass x-ray binaries \citep{3pc} as well as discover previously unknown variable sources \cite{2019ApJ...884...93C,2022ApJ...935....2C} and probe the emission mechanisms of high-mass gamma-ray binaries \citep[e.g.,][]{2025ApJ...986..181L}, gamma-ray pulsars \citep[e.g.,][]{2025ApJ...988..200L} and microquasars \cite[e.g.,][]{2009Sci...326.1512F}.

\texttt{FermiTools} is the official software developed by the SLAC National Accelerator Laboratory and the Fermi--LAT Collaboration for reducing and analyzing Fermi--LAT data. It includes command-line tools and modules for processing, filtering, and modeling LAT data, supporting tasks such as event selection, and likelihood modeling\footnote{For more detailed information and tutorials, please see the FermiTools documentation at: \url{https://fermi.gsfc.nasa.gov/ssc/data/analysis/documentation}.}. Built on top of FermiTools, Fermipy, is a Python-based high-level analysis package that streamlines much of the tools into easy usage in python. Fermipy \citep{fermipy2017} provides a user-friendly interface for performing source detection, spectral fitting, and generating maps and light curves\footnote{For documentation and tutorials, visit: \url{https://fermipy.readthedocs.io/en/latest}.}. Similarly, there have been user-contributed routines written in various languages \citep[e.g., like\_bphase.pl\footnote{See FermiTools User Contributed Software: \url{https://fermi.gsfc.nasa.gov/ssc/data/analysis/user}.} and EasyFermi;][]{2022A&C....4000609D} to further simplify the reduction and analysis steps\footnote{For documentation, visit: \url{https://easyfermi.readthedocs.io/en/latest}.}. 

There is, however, a major issue that one can run into, and that is that many analyses can be quite computationally expensive. A phase-resolved (e.g., useful for a spectral changes as a function of pulsar spin or binary orbit) likelihood analysis requires a computationally expensive requirement to generate livetime cubes, exposures maps as the time to complete the analysis is directly related to the number of phase bins.

To avoid this, many users have turned to high-performance computing and multi-threading on clustered networks where computational resources are abundant and can easily be split to cater to a high volume of analyses of phase bins (among other things). In this paper, a novel and steamlined procedure to undertake phase-resolved likelihood analysis, \texttt{FermiPhased}, is introduced and its uses explored. In Section \ref{sec:2}, the basics of the procedure are detailed for the designed GUI. In Section \ref{sec:3}, case uses of FermiPhased are discussed, including the ``standard'' test case, a ``fixed count'' case (useful for pulsar analysis), and a ``joint phase-resolved'' likelihood fit (useful for sources that exhibit similar behavior like flaring states). In Section \ref{sec:4}, the application of FermiPhased for clusters is discussed and in Section \ref{sec:5}, we discuss the installation and dependencies required to use FermiPhased. In Section \ref{sec:6}, we make our concluding remarks.

\begin{section}{FermiPhased}
\label{sec:2}
We introduce the graphical application \texttt{FermiPhased}\footnote{More information can be found on Github: \url{https://github.com/aplangrangian/FermiPhased}.}, released under the Apache License, Version 2.0 (Apache Software Foundation 2004), that is designed to streamline the generation of Fermi--LAT analysis scripts for phase-resolved studies. It offers a user-friendly interface for inputting source parameters—such as right ascension, declination, period, reference epoch (T$_0$), energy range, and time intervals—along with the option to load or save these configurations via JSON files (see Figure~\ref{fig:fig1}). With labeled input fields and a structured layout, this automated script generator provides shell scripts to call and run FermiTools, create configuration files (.\texttt{YAML}) for each phase, and a basic FermiPy analysis script. \texttt{FermiPhased} can also be used from the command line, although the graphical interface tool-kit PyQt5 is the focus of this paper. 

The source code should be slightly modified for the use of remote clusters, for example, using Rivest-Shamir-Adleman (RSA) keys and log-in info with the user's remote server. The user can browse for spacecraft and event files, specify output directories, and choose to generate a a series of bash scripts ready for execution either on their local machine or a remote cluster. The  automatically generated configuration files and analysis scripts to fit the Fermi sources use the user-specified inputs and save outputs  such as phase bins, integrated fluxes, spectral parameters, and test statistics.
\begin{figure*}
    \centering
    \includegraphics[width=1\linewidth]{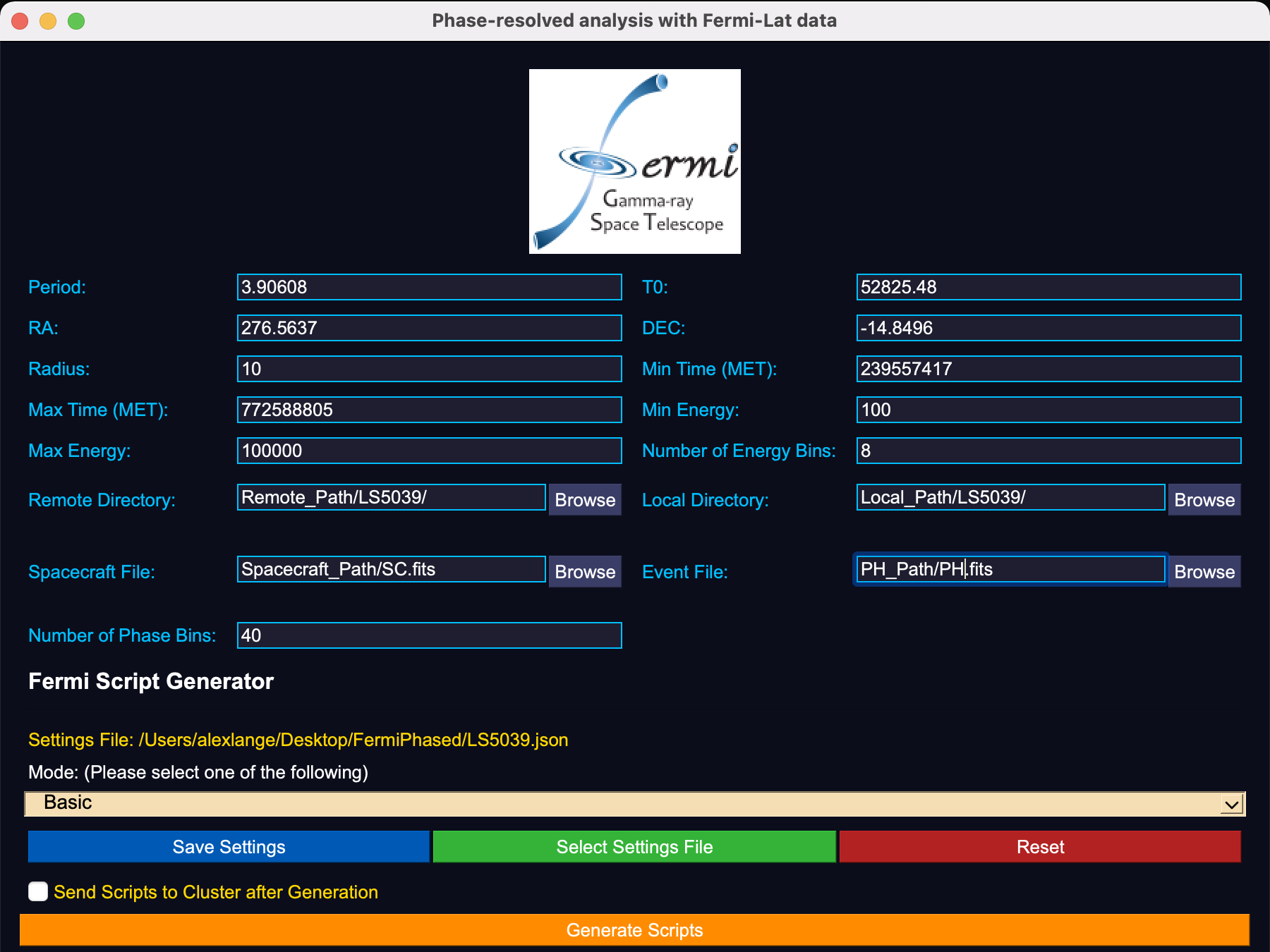}
    \caption{Graphical user interface (GUI) for FermiPhased, a script generator for phase-resolved analysis of Fermi-LAT data. The user can specify key parameters such as source position, energy and time ranges, binning configuration, and file paths. FermiPhased supports multiple modes of phase binning (e.g., Standard Phase-Resolved Analysis, Adaptive (Fixed Counts) Phase-Resolved Analysis, Joint Phase-Resolved Analysis) and allows script export with optional automatic cluster submission for batch processing. Parameters and settings can be saved and loaded here.}
    \label{fig:fig1}
\end{figure*}

The scripts call the FermiTools \texttt{gtmktime}, \texttt{gtselect}, \texttt{gtbin}, and \texttt{gtltcube} after accounting for the user inputs. Upon executing these commands, the script performs the likelihood analysis (using FermiPy's \texttt{gtlike} wrapper) in a generated python file and saves any useful scientific products into a .csv file. For this script to work, the relevant .fits files (photon files and spacecraft files) must be downloaded from the Fermi--LAT data server\footnote{More information and caveats at \url{https://fermi.gsfc.nasa.gov/ssc/data/access}}. Lastly, all analyses are scripted to use the LAT Pass 8 (P8R3\_V3) data \citep{pass82018}.

Executing the scripts generates a subdirectory, corresponding to each phase bin, within the working directory and starts with \texttt{gtmktime}. The given \texttt{gtmktime} command applies a time filter to Fermi-LAT event data to extract time intervals corresponding to a specific phase bin in a periodic signal (e.g., from a binary system). The filter selects events where both the start and stop times fall within a certain phase window ($\phi$), defined by functions that represent the phase selection. The time is converted from Mission Elapsed Time (MET; in seconds) to Modified Julian Date (MJD), shifted by a reference epoch, $T_0$, and folded on the orbital or rotational period, P (days), to isolate the desired phase slice (of width corresponding to 1/number of bins, $n$). For example, for an analysis with 12 phase bins (n=12), the range of phases for the first phase bin is $\phi[i=0] = [0.9583,0.04167]$, where $i = 0, ..., n-1$. An illustration can be seen in Figure~\ref{fig:fig2}. Only good quality data are selected with the keywords ``DATA\_QUAL $> 0$'' and ``LAT\_CONFIG == 1''.
\begin{figure}
    \centering
    \includegraphics[width=0.9\linewidth]{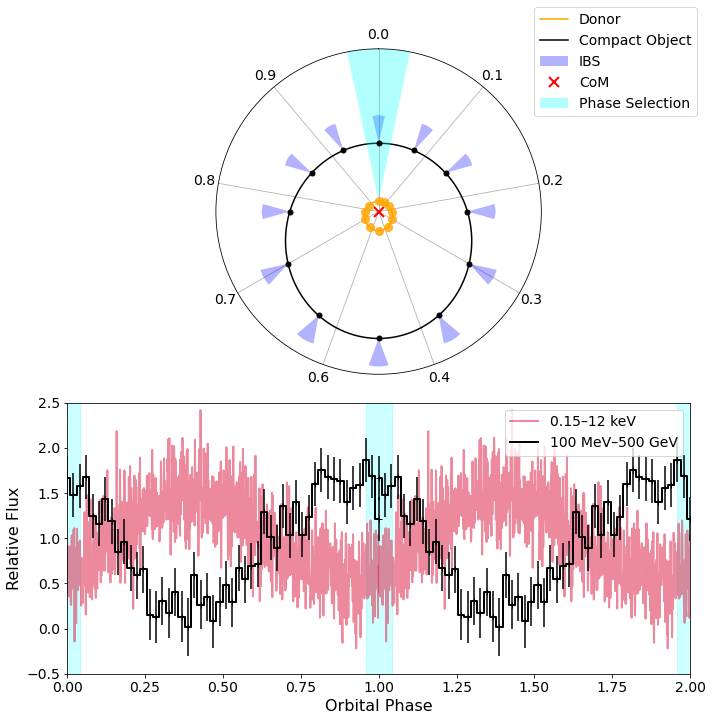}
    \caption{Phase-folded light curve and polar histogram generated by FermiPhased for a selected region of interest in phase space.  The blue-shaded region indicates a user-defined phase interval selected for data filtering and analysis based on user inputs. {\bf Top:} A visualization of the phaseogram along with a schematic of an example binary. {\bf Bottom:} The same phaseogram highlighting modulation features and selected phase cuts. Two orbital cycles are given for clarity.}
    \label{fig:fig2}
\end{figure}
The phase condition applied to both START and STOP times is simply:

\begin{equation}
    \left|\,\phi - \frac{i}{n}\,\right| < \frac{1}{2n},
\end{equation} where is the orbital phase at each event:
\begin{equation}
    \phi = \left(\frac{\texttt{START\,(MJD)} - T_0}{P}\right) \bmod 1.
\end{equation}

After filtering the event times to the corresponding phases, \texttt{gtselect} further filters according to other user-defined selections (energy range, angle separation from the source right ascension (RA) and declination (DEC), overall time range, and other standard reduction choices) and saves the filtered events to a .fits file. \texttt{gtbin} and \texttt{gtltcube} similarly use the user-defined selections along with the standard reduction options to generate counts cube and livetime cube files required for a likelihood analysis. All data products are saved within their designated subdirectory within the working directory (either locally or on a cluster).

The script then generates a configuration file for each subdirectory, and executes a python script (with relevant user inputs) to analyze the source of interest with \texttt{gtlike} and generates an SED for the source. The most up-to-date galactic and isotropic diffuse models\footnote{Both of which can be found here \url{https://fermi.gsfc.nasa.gov/ssc/data/access/lat/BackgroundModels.html}. For our purposes we use gll\_iem\_v07.fits and iso\_P8R3\_SOURCE\_V3\_v1.txt to model the galactic diffuse and isotropic diffuse backgrounds.} and any extended source model files\footnote{\url{https://fermi.gsfc.nasa.gov/ssc/data/access/lat/14yr_catalog}} (.xml or .fits) are required for the generated region of interest (ROI).
\section{Analysis Modes}
\label{sec:3}
A central feature of \texttt{FermiPhased} is its support for three different analysis modes, each designed to accommodate different scientific goals. Currently, there are three modes, \textbf{Standard Mode}, in which one can run a standard phase-resolved analysis, \textbf{Constant Counts Mode}, in which one can create dynamically spaced phase bins so that each phase bin has the same number of counts (maintaining similar S/N) , and \textbf{Joint Phase-Resolved Analysis}, where one can choose to analyze multiple time ranges in a joint-likelihood phase-resolved analysis.

\subsection{Standard Phase-Resolved Analysis}\label{sec:standard}
In the \textbf{Standard Mode}, the user defines a single time interval along with the number of desired phase bins. The application then divides the orbit into equal phase intervals and generates scripts for each bin, filtering events based on either their orbital or rotational phase. This approach is ideal for sources that exhibit periodic signal in other wavelengths that have not yet been explored in the gamma-ray regime. Figure~\ref{fig:fig1} shows an example analysis set-up for high-mass gamma-ray binary LS 5039.

As an example, we choose a 17  year observational period of the gamma-ray binary LS 5039 (4FGL J1826.2-1450). LS 5039 is a high-mass gamma-ray binary (HMGB) consisting of a compact object, that is likely to be a neutron star and a massive O-type companion star \citep{casares_possible_2005}. LS 5039 is relatively bright in the  $\gamma-$ray sky and shows modulation on its orbital period. With a short orbital of 3.90608\,days \citep{casares_possible_2005,aragona_orbits_2009,2009ApJ...706L..56A,volkov_nustar_2021}, it is a good test source to study the modulation of gamma-ray emission as a function of orbital phase.

We choose all events within 10 degrees of LS 5039 (RA, Dec) = (276\fdg5637,-14\fdg8496) within the time range of photons between 239557417 and 772588805 MET (2008 August 04 to 2025 June 26). Additionally, we only consider events more energetic than 100\,MeV and less than 100\,GeV. We fold the events on the orbital period and divide into 40 phase bins of equal width ($\delta \phi = 0.025$). These data selections are shown in Figure~\ref{fig:fig1}.

We show the results of our phase-resolved analysis likelihood in Figure~\ref{fig:fig2.5}. This is the highest phase resolution conducted on LS 5039 \citep{2009ApJ...706L..56A,2012ApJ...749...54H,2019RAA....19..180C,2021ApJ...917...90Y}. We choose to model LS 5039 with the 4FGL-DR4 catalog model, the LogParabola spectral model. The best-fit spectral parameters $\alpha$ and $\beta$ show modest variation with statistically interesting features around phase $\phi\approx0.35$ and $\phi\approx0.70$ (which corresponds to the inferior conjunction), showing a brief increase in hardness in the spectral slope. LS 5039's spectrum changes its curvature ($\beta$) slightly throughout the orbit, with the most curvature found near the periastron ($\phi=0.0$). These results could be useful for phase-resolved multi-wavelength SED modeling \citep[e.g., 4FGL J1405.1-6119;][]{2025ApJ...986..181L}, though a full analysis of LS 5039 is beyond the scope of this paper.
\begin{figure}[h]
    \centering
    \includegraphics[width=\linewidth]{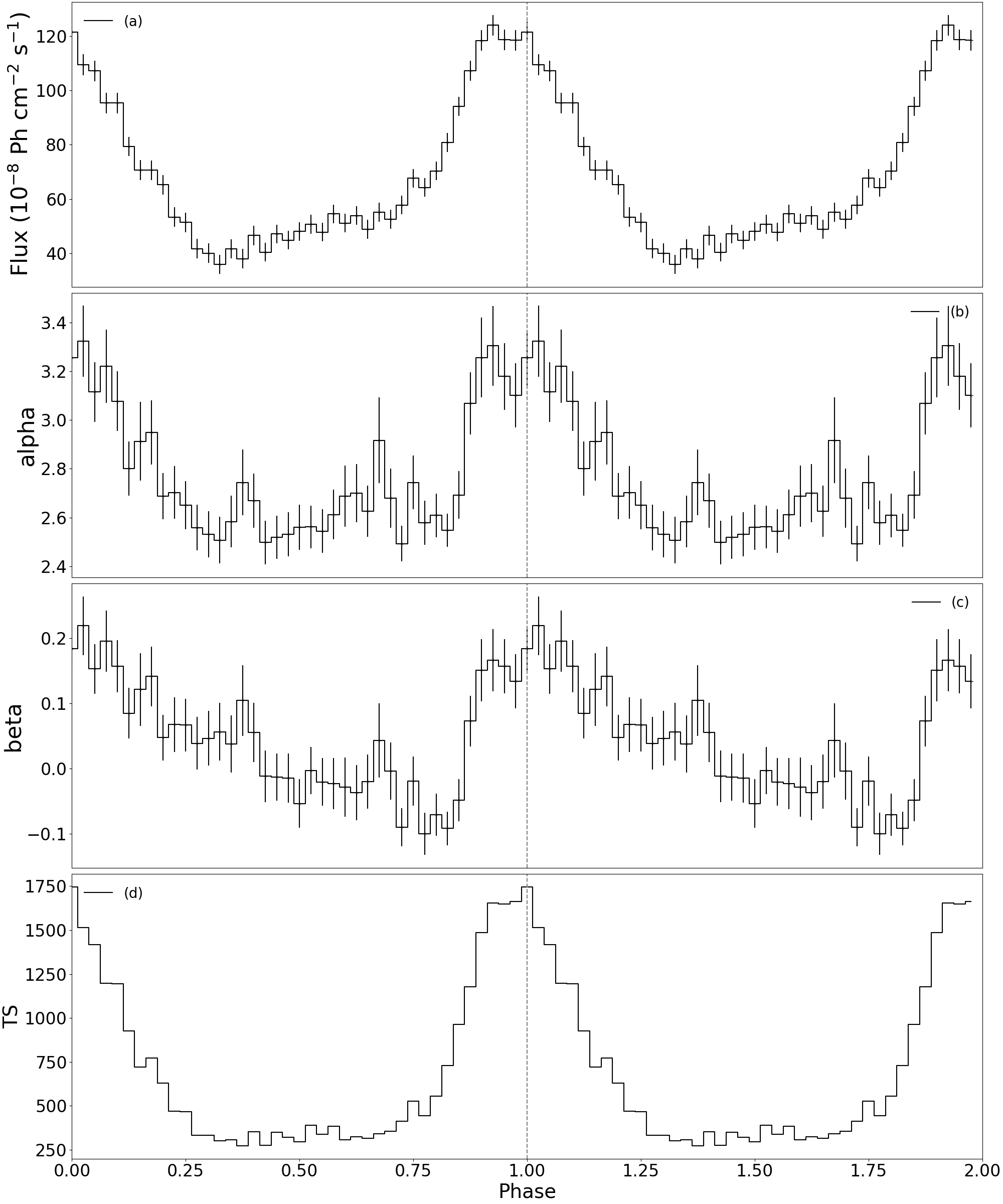}
    \caption{Phase-resolved analysis of LS 5039 from nearly 17 years of Fermi-LAT data. The 0.1--100\,GeV integrated flux is shown in the top panel, spectral parameters $\alpha$ and $\beta$ are shown in the middle two panels. The TS for LS 5039 is shown in the bottom panel. Two orbital cycles are shown for clarity.}
    \label{fig:fig2.5}
\end{figure}
    
\subsection{Adaptive (Fixed Counts) Phase-Resolved Analysis}
The \textbf{Adaptive (Fixed Counts) Mode} takes a different approach, allowing users to define a desired number of counts per phase bin rather than a fixed number of bins. This approach is particularly useful for the phase-resolved spectral analysis of variable sources with dramatic changes in flux, such as in the case of a pulsar. With a timing model (e.g., Tempo2, PINT\footnote{See PINT's documentation at: \\ \url{https://github.com/nanograv/PINT}.\label{fn:fn9}}) to assign a phase to each photon, \texttt{FermiPhased} will then dynamically create the phase intervals for each bin.
\begin{figure}[h]
    \centering
    \includegraphics[width=1\linewidth]{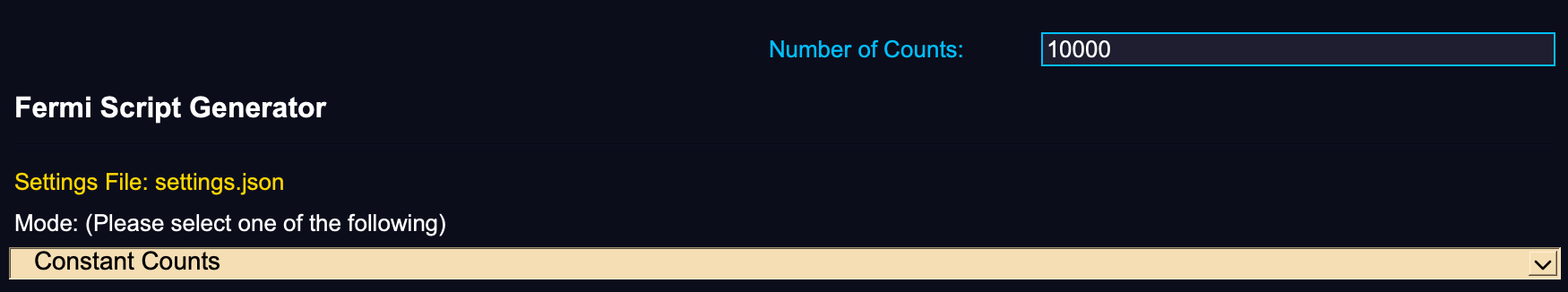}
    \caption{Zoomed-in image of FermiPhased interface in “Adaptive (Fixed Counts)” mode, where the user specifies the desired number of counts per phase bin.}
    \label{fig:fig3}
\end{figure}

For instance, the pulsar timing tool, PINT \citep[PINT Is Not Tempo3;][]{2021ApJ...911...45L,2024ApJ...971..150S}, converts photon arrival times into rotational phase values based on a timing model of a pulsar or binary (as does Tempo2; \citet{2006MNRAS.369..655H}). Photon times are also barycentered from MET to Barycentric Dynamical Time (TDB
). Then, using a pulsar ephemeris, either created by PINT or taken from another observatory (e.g., Crab pulsar ephemeris from Jodrell Bank Observatory\footnote{More information can be found at: \\ \url{https://www.jb.man.ac.uk/pulsar/crab.html}.}), PINT computes the phase of the pulsar’s rotation at each photon’s arrival time. These phases are then used for pulse profile studies, folding, and phase-resolved analyses like the binning described above.

After filtering the events by energy, the photon pulse phases are extracted and sorted. The sorted phases are divided into bins containing a fixed number of counts (e.g., Figure~\ref{fig:fig3} with 30,000 photons per bin), allowing the bin widths to vary depending on the local photon density. 

We use the well-studied Vela pulsar to test the adaptive phase binning and analysis mode. The Vela pulsar (4FGL J0835.3-4510) is the brightest steady GeV $\gamma-$ray source in the sky and has served as a quintessential target for pulsar studies and modeling with Fermi-LAT \citep{abdo_fermi_2009,abdo_psr_2010,abdo_pwn_2010,2025ApJ...988..200L} along with the Crab pulsar. The Vela pulsar, (RA, Dec) = (128\fdg8370, -45\fdg1781), has a spin period of $\sim89$\,ms. Its clear double-peaked $\gamma-$ray pulse profile is well modeled and recovered using phase-folding with \texttt{FermiPhased} inputs, which select events from an ROI of 15\degree \,radius around the pulsar, restrict energies to $>60$ MeV to minimize the contribution of background sources due to Fermi--LAT's larger PSF at low energies.

In Figure~\ref{fig:fig4}, we display how phase bins are created as a function of pulse phase, specific to the Vela pulsar. Narrower bins correspond to pulse phases near the main and secondary peaks while wider bins appear in regions of lower flux (during inter-pulse or off pulse phases). For detailed results and analysis of the Vela pulsar as analyzed from 2008 August to
2021 October using FermiPhased, see \cite{2025ApJ...988..200L}.

\begin{figure}
    \centering
    \includegraphics[width=0.75\linewidth]{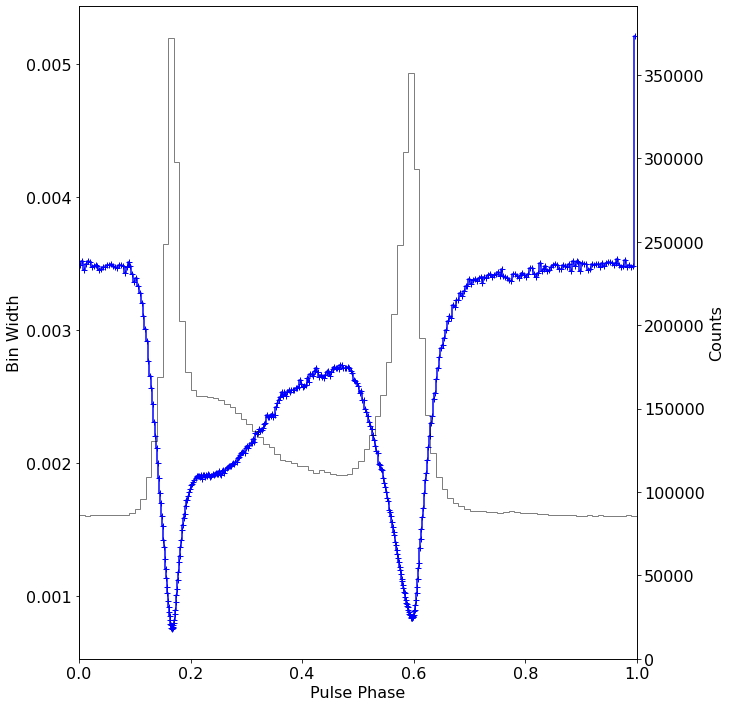}
    \caption{Widths of phase bins as a function of pulse phase in “Adaptive (Fixed Counts)” mode for the Vela pulsar. The bin widths (blue) vary inversely with count rate (grey).  }
    \label{fig:fig4}
\end{figure}
\subsection{Joint Phase-Resolved Analysis}
\textbf{Joint Phase-Resolved Analysis} enables the user to specify multiple time segments, each with their own period and T$_0$ values, ideal for sources whose timing properties evolve over time. For each phase bin, the generated scripts apply time and phase selections across the specified epochs. This mode is especially powerful for studying long-term changes in emission properties or for combining data across multiple outbursts or orbital cycles, all while preserving phase coherence within each segment (assuming accurate ephemerides). 

The $\gamma-$ray source 4FGL J1702.7-5655 (hereafter J1702) has been identified as a promising redback–type binary candidate. Fermi--LAT observations reveal a strong modulation at an orbital period of $\sim0.2438033$ days, and the folded $\gamma-$ray light curve shows a narrow eclipse feature, consistent with the companion periodically occulting the high-energy site of emission. The orbital modulation behavior changed around MJD 56,500, from a simpler eclipse-only profile to a more sinusoidal modulation. More significantly, the eclipse is reported to shift $\delta\phi \approx 0.05$ in phase, indicative of an evolving geometry in the intrabinary shock region \citep{2022ApJ...935....2C}.

As such, this system offers a compelling case for applying phase and time-selection tools to jointly study the different epochs in the $\gamma-$ray light curves. 

Figure~\ref{fig:fig5} shows an example input for J1702, (RA, Dec) = (255\fdg6945, -56\fdg9283) \citep{4fgl2020}, and a comma-separated time range corresponding to Figure~\ref{fig:fig7}'s top panel. This time range is based off periods of gamma-ray modulation of the super-orbital period of J1702.7-5655 (hereafter J1702) as presented in \cite{2022ApJ...935....2C}. 
Similar to the two previous cases, we use a ROI of 10\degree\ around J1702 and only consider events above 100\,MeV and less than 500\,GeV. 
\begin{figure}[h]
    \centering
    \includegraphics[width=1\linewidth]{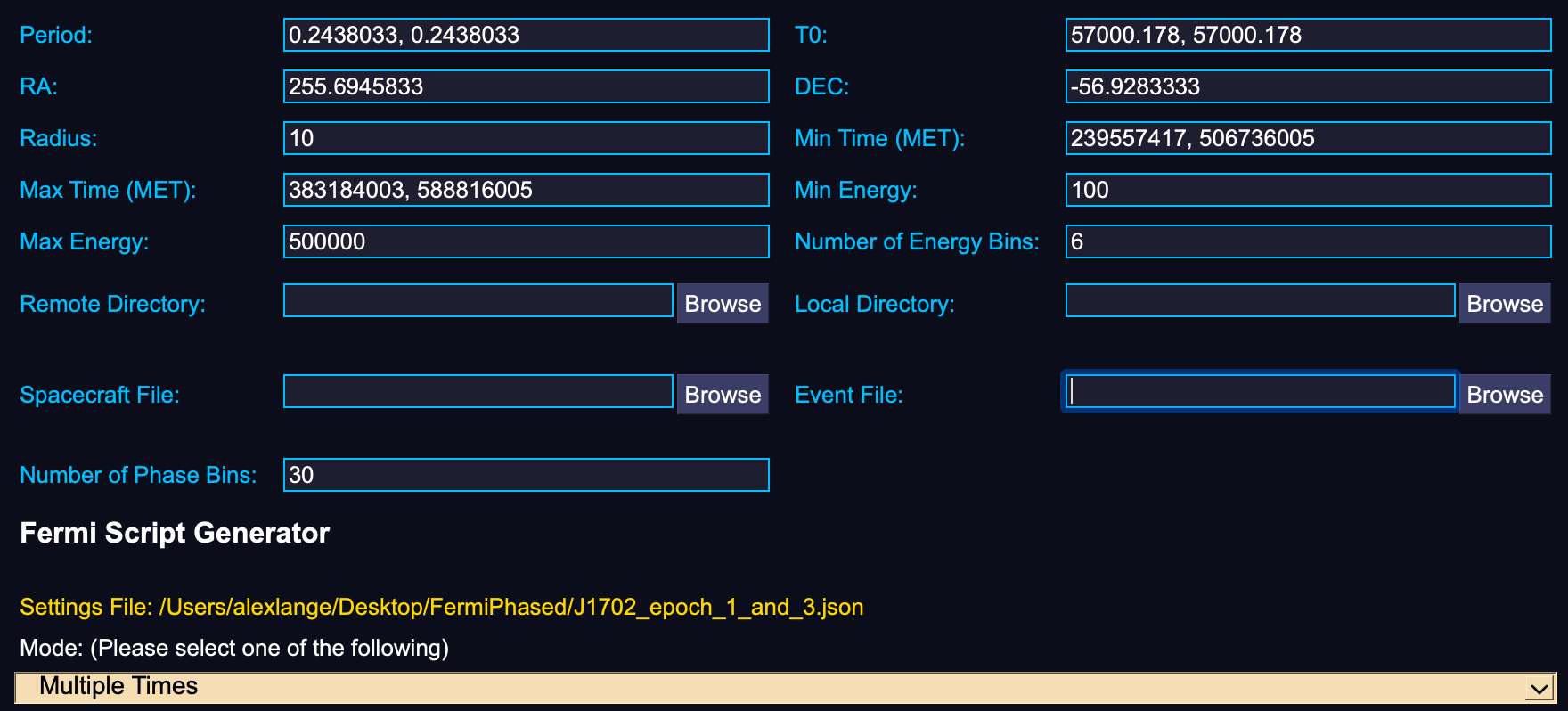}
    \caption{Zoomed-in image of FermiPhased interface in “Multiple Times” mode, which allows users to define distinct analysis intervals with different periods, time ranges, periods and reference epochs. The input parameters (e.g., RA, Dec, etc) are for 4FGL J1702.7-5655.}
    \label{fig:fig5}
\end{figure}

\begin{figure}[h]
    \centering
    \includegraphics[width=1\linewidth]{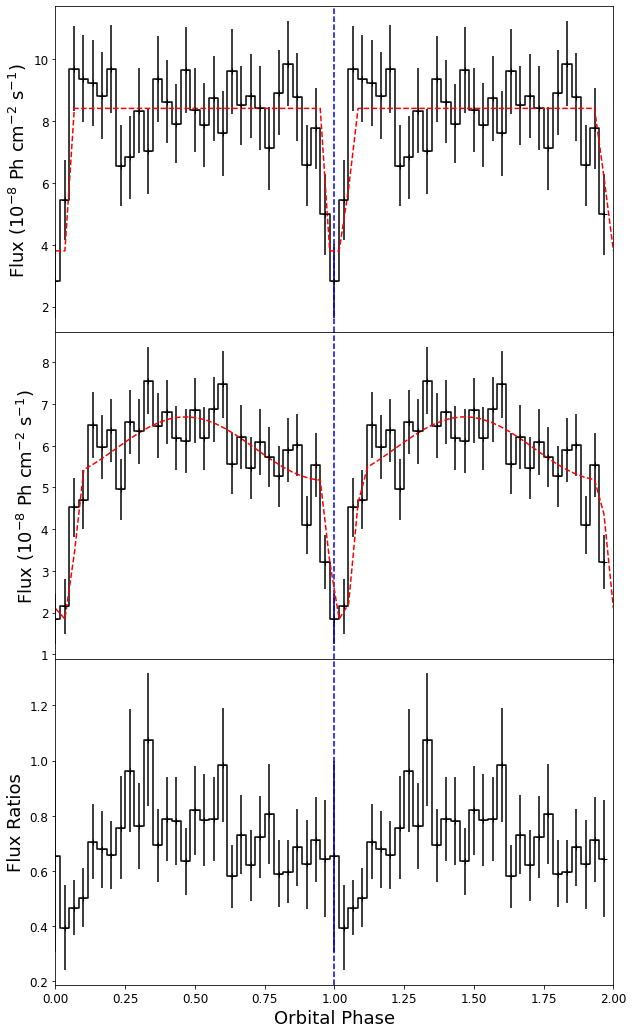}
    \caption{Joint likelihood analysis of two sets of time ranges of candidate redback 4FGL J1702.7-5655. Time ranges were selected by the detection of modulation of flux at the super-orbital period  associated with the source using Fermi-LAT aperture fluxes \citep{2022ApJ...935....2C}. {\bf Top:} The joint likelihood analysis of all selected Fermi events between MJD 54,682 and MJD 56,345 as well as all events between MJD 57,775 and MJD 58,725. {\bf Middle:} The same but for all Fermi events from MJD 56,345 and MJD 57,775 as well as between MJD 58,725 and MJD 60,488. {\bf Bottom:} The flux ratios of the top and middle panels.}
    \label{fig:fig7}
\end{figure}

Together, these modes provide an adaptable framework for constructing time- and phase-resolved analyses of periodic or variable gamma-ray sources. By automating the mundane and error-prone task of writing bash scripts for each phase interval, the tool empowers users to focus on interpreting scientific results. Whether the goal is standard folding, adaptive binning, or joint phase-folded analysis of multiple epochs, this interface ensures that setup and script generation are fast, consistent, and reproducible.

\end{section}
\section{Cluster Compatibility}
\label{sec:4}
Phase-resolved analysis of $\gamma-$ray sources using Fermi-LAT data is inherently computationally intensive. For each phase bin, one must filter the data, compute the exposure, generate models, fit spectra, which require invoking Fermitools (and Fermipy) commands across multiple energy bands and time intervals. When expanded across dozens of phase bins, energy ranges, and possibly multiple epochs or parameter sets, the processing load becomes too expensive for a local machine to handle. 

While \texttt{FermiPhased} can  save the analysis scripts locally, it can also address this by creating and sending scripts to a high-performance computing cluster using \texttt{Slurm}\footnote{See Slurm documentation: \\ \hyperlink{https://slurm.schedmd.com/documentation.html}{https://slurm.schedmd.com/documentation.html}.} job scheduling to be sent to and executed remotely on remote computing cluster (if the appropriate checkbox is checked, see Figure~\ref{fig:fig1}). 

The analyses of the phase bins are parallelized across different nodes, enabling substantial speedups and resource isolation. To do so, the shell scripts contain Slurm commands (commands to interact with the slurm job scheduler) to direct the analysis to run on a specific partition (\#SBATCH -p partition\_name), on a requested number of compute nodes (\#SBATCH -N number\_of\_nodes; likely 1) for some maximum time (\#SBATCH -t time\_limit). The analysis is lastly set to run inside the working directory  (\#SBATCH -D your\_working\_directory), ensuring that all input and output files are handled correctly. The scripts activate the Fermi environment as set environment variables used for the FermiTools.

This parallelization ensures consistent runtime performance and prevents memory bottlenecks or crashes associated with heavy workloads. For instance, the average working time for the standard analysis mode job in Section~\ref{sec:standard} was $\sim2.5$\,hours to complete. While one may use these scripts on a local machine (with minimal threading possibilities $\sim8$ cores), one may also be forced to conduct their analyses in sequential order, calling the analysis of each phase bin after another (as determined by the number of cores), resulting in long computation times. Integration with SLURM provides the ability to analyze many (if not all) phase bins at once as well as job and error/logging management.

\texttt{FermiPhased} automates script generation and, with the help of SSH and SCP (standard secure transfer protocols), transfers the scripts to a cluster for execution. 

\section{Dependencies, Installation and Usage}
\label{sec:5}
The Fermi Script Generator is a Python-based graphical user tool designed to streamline the production of shell scripts and configuration files for phase-resolved analysis of Fermi--LAT data. The code relies on a small set of open-source Python libraries together with the standard Fermi Science Tools environment. The graphical interface is implemented in PyQt5, while numerical routines make use of \texttt{NumPy}. Configuration files are written in YAML using the \texttt{PyYAML} library, and cluster file transfers are supported through \texttt{paramiko} and \texttt{scp}. Optional post-processing and visualization routines require \texttt{matplotlib} and \texttt{fermipy}. A functioning installation of the Fermitools (including \texttt{gtselect}, \texttt{gtmktime}, \texttt{gtbin}, and \texttt{gtltcube}) is assumed to be on the local machine or the intended cluster to be used.

Installation is straightforward: all Python dependencies may be installed using \texttt{pip} or \texttt{conda} (i.e.,  with the command ``pip install pyqt5'' in the command line in the conda environment with an installation of FermiTools and FermiPy). A typical setup involves creating a dedicated conda environment that includes the Fermitools and FermiPy distribution and then installing the additional Python packages (\texttt{pyqt5}, \texttt{numpy}, \texttt{pyyaml}, \texttt{paramiko}, \texttt{json}, and \texttt{scp}). Additionally, to assign pulse phases with PINT, your working environment requires an installation of PINT (``pip install pint-pulsar''\footnote{See footnote \ref{fn:fn9}.}). The user should also ensure that the remote computing cluster is configured with slurm for job scheduling and that the appropriate Fermi calibration photon dataset and diffuse models are accessible. Once the Python environment is in place, the script generator itself can be obtained by cloning the repository or copying the source directory to a local working location. The tool is launched simply with \texttt{python fermi\_script\_generator.py}.

\section{Conclusion}
\label{sec:6}
\texttt{FermiPhased} enables users to generate reproducible, batch-submitted analyses on a high-performance cluster or on a local machine. By encapsulating both the low-level command construction and the bookkeeping of configuration files, the tool minimizes human error, and generates an intuitive analysis pipeline.

Upon startup, the user is presented with a graphical interface where source parameters and file paths  can be entered. The interface allows switching between different analysis modes: fixed phase bins, constant counts, or multiple time intervals. For each configuration, the generator produces phase-resolved analysis scripts (\texttt{.sh}) together with YAML configuration files describing the analysis setup. The user may optionally enable automatic transfer of these files to a computing cluster. A final analysis  script is also produced to finish the likelihood analysis, allow for quick customization and plotting once all slurm jobs have completed. A simple SLURM command allows for email updates.

Overall, this application serves as an efficient and user-friendly front-end for researchers working with Fermi LAT data. It simplifies complex script creation, enables reproducibility through saved configurations, and integrates cluster submission—all within a self-contained GUI. Its flexibility and automation capabilities make it a useful tool for both professional high-energy astrophysicists or students studying pulsars, binaries, or other periodic sources with Fermi--LAT.

\acknowledgments
The authors also thank Dr. Robin H.D. Corbet and Dr. Oleg Kargaltsev for useful reviews and comments as well as mentorship using FermiTools.

\pagebreak
\bibliographystyle{aa}
\bibliography{main.bib}
\end{document}